\documentclass[fleqn,usenatbib]{mnras}

\usepackage{newtxtext,newtxmath}

\usepackage[T1]{fontenc}
\usepackage{ae,aecompl}

\usepackage{graphicx}	
\usepackage{amsmath}	
\usepackage{tikz}
\usepackage{orcidlink}
\usepackage{hyperref}

\title[AARTFAAC-12 characterization]{Characterization of the AARTFAAC-12 aperture array: radio source counts at 42 and 61 MHz}

\author[A.~Shulevski et al.]{
A.~Shulevski,$^{1,2,3}$\thanks{E-mail: shulevski@strw.leidenuniv.nl} \orcidlink{0000-0002-1827-0469} T.~M.~O.~Franzen,$^{2}$ W.~L.~Williams,$^{3}$ T.~Vernstrom,$^{6}$ 
\newauthor B.~K.~Gehlot,$^{4,5}$ \orcidlink{0000-0002-3240-9228} M.~Kuiack,$^{1}$ R.~A.~M.~J.~Wijers,$^{1}$
\\
$^{1}$Anton Pannekoek Institute, University of Amsterdam, Postbus 94249, 1090GE Amsterdam, The Netherlands\\
$^{2}$ASTRON: the Netherlands Institute for Radio Astronomy, PO Box 2, 7990 AA, Dwingeloo, The Netherlands\\
${^3}$Leiden Observatory, Leiden University, PO Box 9513, NL-2300 RA
Leiden, The Netherlands\\
$^{4}$Kapteyn Astronomical Institute, University of Groningen, PO Box 800, 9700AV Groningen, The Netherlands\\
$^{5}$School of Earth and Space Exploration, Arizona State University, Tempe, AZ, USA\\
$^{6}$CSIRO Astronomy \& Space Science, Kensington Perth 6151, Australia\\
}

\date{Accepted XXX. Received YYY; in original form ZZZ}
\pubyear{2022}
\hypersetup{draft} 
\begin{document}
\label{firstpage}
\pagerange{\pageref{firstpage}--\pageref{lastpage}}
\maketitle

\begin{abstract}
Dense aperture arrays provide key benefits in modern astrophysical research. They are flexible, employing cheap receivers, while relying on the ever more sophisticated compute back-end to deal with the complexities of signal processing required for optimal use. Their advantage is that they offer very large fields of view and are readily scalable to any size, all other things being equal. Since they represent "software telescopes", the science cases these arrays can be applied to are quite broad.
Here, we describe the calibration and performance of the AARTFAAC-12 instrument, which is composed of the twelve centrally located stations of the LOFAR array. We go into the details of the data acquisition and pre-processing, we describe the newly developed calibration pipeline as well as the noise properties of the resulting images and present radio source counts at 41.7 MHz and 61 MHz.
We find that AARTFAAC-12 is confusion limited at 0.9 Jy/PSF at 61 MHz with a PSF size of $17\arcmin\times11\arcmin$ and that the normalized source counts agree with the scaled VLSSr and 6C survey counts. The median spectral index of the sources between the two frequencies we observed at is -0.78. Further, we have used the derived source counts to estimate any excess cosmic radio background, and we do not find evidence for it at our observing frequencies compared to published literature values.
\end{abstract}

\begin{keywords}
instrumentation: interferometers -- methods: observational -- techniques: interferometric -- radio continuum: general
\end{keywords}



\section{Introduction}
\label{sec:intro}

Recent developments in the area of signal processing in astronomy as well as the declining cost of high-performance (computing) hardware has enabled the advent of (dense) aperture array instruments in radio astronomy \cite[][LOFAR, MWA, HERA, LWA, NenuFar, OVRO-LWA]{vanHaarlem2013, Beardsley2019, deBoer2017, Taylor2014, Zarka2012, Hallinan2015}. This is especially true for low frequencies (up to a few GHz). At the lowest frequencies (tens to hundreds of MHz), the arrays use simple dipoles as receiving elements which synthesize an aperture (hence aperture arrays) to various degrees, by sampling the electric field of the incoming electromagnetic radiation and employ simple receivers for initial signal conditioning in the field. These systems usually have very large fields of view (FoV) up to $ 2\pi \, \mathrm{sr}$, depending on the antenna pattern of the individual receiving element; it can be modified by combining elements together (beam-forming) if so desired, diminishing the FoV, but gaining in sensitivity.

The LOw Frequency ARray \cite[LOFAR;][]{vanHaarlem2013} telescope uses dipoles sensitive to frequency ranges of 10-90 MHz (low band, LBA) and 110-240 MHz (high band, HBA) grouped into stations which are then correlated with each other or their signals added (depending on the science case). In another incarnation of (part of) the same instrument, demonstrating its modular design, we have taken the LBA dipoles of the central twelve stations, extracted the dipole signals out of the LOFAR signal path before station beam forming and correlated them with each other. Thus, we have obtained a dense array, with a very large FoV, the Amsterdam-Astron Transient Facility And Analysis Center (AARTFAAC-12, henceforth A12).

A12 is able to observe commensally with LOFAR, whenever it is observing in LBA mode (we usually use the {\tt LBA\_OUTER} mode in which a ring of 48 outermost LBA dipoles in each station is active). We use a separate correlator to produce visibilities on baselines between each of the dipoles comprising the 12 core stations, and can form up to 16 sub-bands (SBs), each with a bandwidth of 195.3 kHz across the LBA band. These SBs can be placed at any desired frequency in various configurations (continuous or dispersed). Each SB can be subdivided into up to 64 channels, but the correlator performance starts to deteriorate if a larger number of channels per SB are recorded. For more details on the system configuration and signal path, we refer the reader to \cite{Prasad2016}.

The scientific use cases of the A12 system are varied; its sensitivity to diffuse extended emission renders it very suitable for Galactic science. Its large FoV and the low frequencies at which it operates make it a powerful tool for space weather studies \citep{Mevius2016} as well as measuring the power spectrum of Cosmic Dawn \citep[z$\sim$18;][]{Gehlot2020}. Since the instrument images the whole sky at high temporal cadence (one second), it is also  suitable for radio transient detection, enabling an order of magnitude improvement in sensitivity compared to its predecessor, AARTFAAC-6 \citep{Kuiack2019}.

To utilize the instrument properly in imaging mode, which is its main mode of operation, we need to characterize its systematics, and assess its performance as well as that of the associated data processing pipeline. In this work, we will describe the A12 calibration and imaging results as well as image noise properties using standard LOFAR processing tools (Section \ref{sec:reduction}) and go on to derive source counts using a typical A12 data set (Section \ref{sec:res}). We discuss our results in Section \ref{sec:disc} and give concluding remarks in Section \ref{sec:conc}.

Even though our source counts are limited to brighter sources, the number of such studies at extremely low frequencies are limited \citep[eg.][]{Lane2014}. Also, since we derive our source counts from a single observation covering a large area of the visible sky at that moment, our results have fewer observational systematics compared with surveys resulting from multiple pointings to cover the same sky area.

\section{Data reduction}
\label{sec:reduction}

We have recorded visibilities for fifteen minutes starting at 01:29:27.5 UT on February 18, 2019. Each SB has 3 channels and we have placed one contiguous group of eight SBs around 41.7 MHz and another 8 SB group around 61 MHz, at the maximum sensitivity of the LBA dipoles. The integration time was one second. A12 observes in drift scan mode, pointing at zenith with a FoV extending over the entire visible sky.

The A12 correlator produces visibilities in near-real time; to facilitate studies of radio transients it was designed to minimize latency using nominal observing parameters. The visibilities are written on disk in a way compatible with the AARTFAAC-6 reduction and imaging pipeline which produces near-real time (up to three seconds lag time) images for transient detection \citep{Prasad2016}. The recording format is a streaming format; to facilitate calibration using LOFAR data reduction packages, we convert the visibilities to measurement set (MS) format using the \textsc{aartfaac2ms} tool \citep{Offringa2015}. During the conversion, we flag the visibilities to mitigate RF interference using the \textsc{AOFlagger} tool \citep{Offringa2012}, flag for bad dipoles (based on auto correlation values) and phase shift the visibilities to the coordinates of zenith at the mid-point of the observing run (in this case $ \alpha: 11^{h}55^{m}01.2^{s} $, $ \delta: +52^{\circ}50\arcmin17.43\arcsec $).

Before initiating the calibration, we concatenate each group of eight individual single-SB measurement sets into a single measurement set comprising 24 frequency channels. The calibration procedure consists of the following steps:

\begin{itemize}
    \item Solving for the direction independent (DI) complex gains of each receiving element per integration time. The beam pattern of the elements is taken into account. We use CasA and CygA as calibrators to set the flux scale, using source models composed of Gaussians as well as points, set on the Scaife and Heald flux scale \citep{Scaife2012}. We find that using the brightest (visible) 3C sources as calibrators does not result in usable calibration solutions; they have too low S/N, even after using solution intervals of a few minutes.
    
    In the DI case, the problem (per baseline) is cast as \citep{Smirnov2011}:
    \begin{equation}
    \label{eq:di}
        V_{\mathrm{pq}} = G_{\mathrm{p}}B_{\mathrm{p}}M_{\mathrm{pq}}(B_{\mathrm{q}})^{H}(G_{\mathrm{q}})^H 
    \end{equation}
    \noindent where $ V_{\mathrm{pq}} $ are the recorded visibilities on the baseline between dipoles $ P $ and $ Q $, $ G $ are the dipole (complex) gain Jones matrices, $ B $ is the dipole beam Jones matrix and $ M $ are the model visibilities on that baseline. $ H $ denotes the Hermitian (complex) transpose operator.
    
    We solve for the amplitudes and phases of all four polarization products using the {\tt full-jones} mode in \textsc{DPPP}, a package which includes various solvers and other tools for MS manipulation, one of the main components of the LOFAR data processing pipeline. The solution interval is set to equal the integration time (one second) and we solve per channel.
    In the {\tt fulljones} solve mode, the gain Jones matrix per dipole is:
    \begin{equation}
    \label{eq:rime}
        G_{\mathrm{p}} = \left( \begin{array}{cc} A_{\mathrm{xx}}e^{\phi_{\mathrm{xx}}} A_{\mathrm{xy}}e^{\phi_{\mathrm{xy}}} \\ A_{\mathrm{yx}}e^{\phi_{\mathrm{yx}}} A_{\mathrm{yy}}e^{\phi_{\mathrm{yy}}} \end{array} \right)
    \end{equation}
    \noindent here $ \mathrm{XX} $, $ \mathrm{XY} $, $ \mathrm{YX} $ and $ \mathrm{YY} $ are the four instrumental (linear) polarizations of the recorded visibilities. $ \mathrm{A} $ and $ \phi $ stand for the amplitude and phase of the gains. 
    \item Direction dependent (DDE) solve (again using the {\tt full-jones} solve mode) is performed using CasA, CygA, TauA and VirA source models (depending on source visibility and elevation for the duration of the observing run) on the Scaife and Heald flux scale. The dipole beam pattern is taken into account, and the solve is done per one second and channel. The DDE solutions are used to subtract the sources solved for from the visibilities. In this case:
    \begin{equation}
    \label{eq:dde}
        V_{\mathrm{pq}} =\sum_{\mathrm{dir}}G_{\mathrm{p}}^{\mathrm{dir}}B_{\mathrm{p}}M_{\mathrm{pq}}(B_{\mathrm{q}})^{H}(G_{\mathrm{q}}^{\mathrm{dir}})^H
    \end{equation}
    \noindent i.e. we have gain solutions per each direction towards a particular calibration source.
\end{itemize}

\begin{figure*}
	\includegraphics[width=\textwidth]{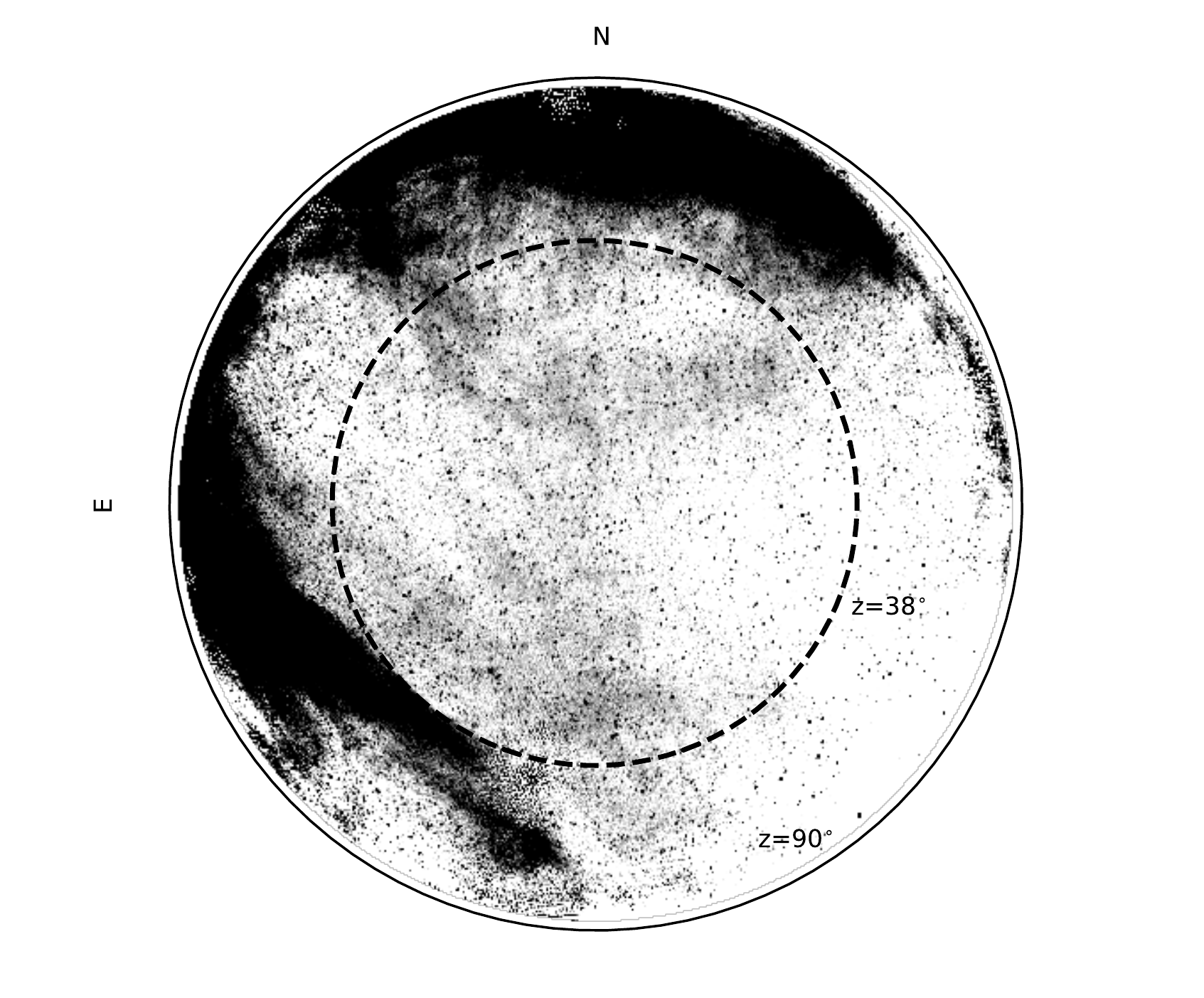}
    \caption{The FoV of the A12 array at 61 MHz (Stokes \textit{I}, no \textit{uv} taper). The image extends to 15$^{\circ}$ elevation, centered on the zenith J2000 coordinates at mid-observation. The Galactic plane is visible along the image edge, with the Galactic spur at bottom left. The PSF is $ 11.4\arcmin \times 16.8\arcmin $ A dashed circle marks the 52$^{\circ}$ elevation limit of the central image cutout used in the source count analysis, and the thin filled line marks the horizon.}
    \label{fig:image:full}
\end{figure*}

We average the calibrated data by a factor of four in frequency (sufficient to avoid bandwidth smearing out to the edge of the FoV) and scale the visibilities with the dipole beam pattern in the direction of the phase center to get intrinsic flux values. Next, we image the averaged data set using WSclean \citep{Offringa2014}. The image scale is 6$\arcmin$ and 3$\arcmin$ per pixel at 41.7 and 61 MHz respectively, and we image the full FoV using robust 0 weights while correcting for the dipole beam during the imaging process. Multi-scale cleaning is used with scales of 0, 10, 30 and 60 pixels. We have specified $ 10^{6} $ clean iterations, letting the clean continue until the stopping criteria determined by the WSclean parameters {\tt auto-mask} and {\tt auto-threshold} were reached. The $ \sigma $ values were set to 3 and 0.3 respectively for these parameters.
We produce Stokes \textit{I} and \textit{V} images without and with a \textit{uv} taper of 10$\lambda$ (corresponding to angular scales larger than six degrees) and removing the diffuse Galactic synchrotron emission to determine the influence it has on our analysis. The images have PSF sizes of $24\arcmin \times 18\arcmin$ and $17\arcmin \times 11\arcmin $ at 41.7 MHz and 61 MHz respectively.
The dipole sensitivity drops sharply near the horizon, which defines our total field of view (FoV). One of the images (non-tapered, Stokes I at 61 MHz), masked starting at the end of the FoV (below an elevation of $ 15^{\circ} $) is shown in Figure \ref{fig:image:full} for reference.

\subsection{Flux scale}
\label{sec:flux_scale}

\begin{figure*}
	\includegraphics[width=0.45\textwidth]{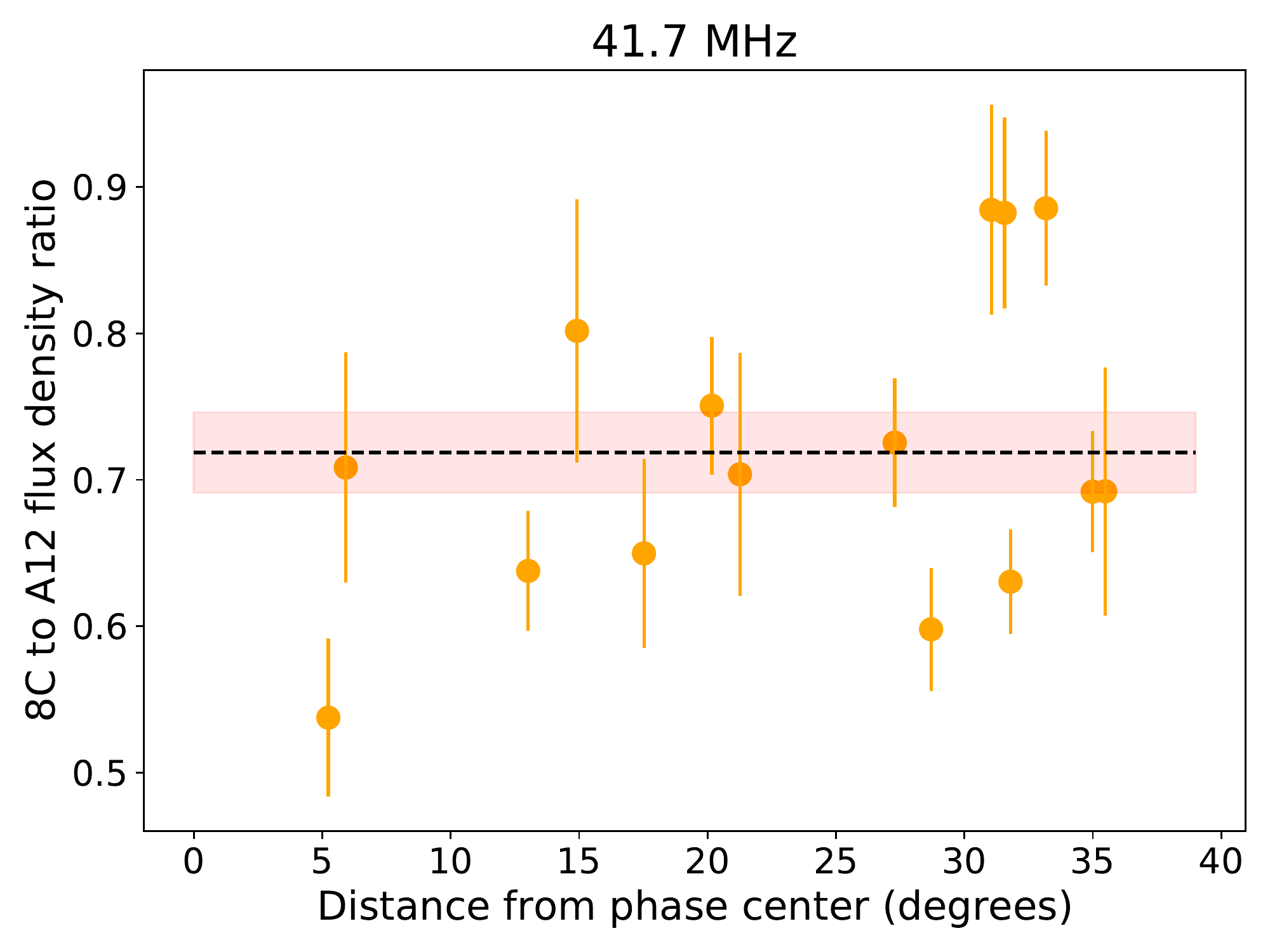}
	\includegraphics[width=0.45\textwidth]{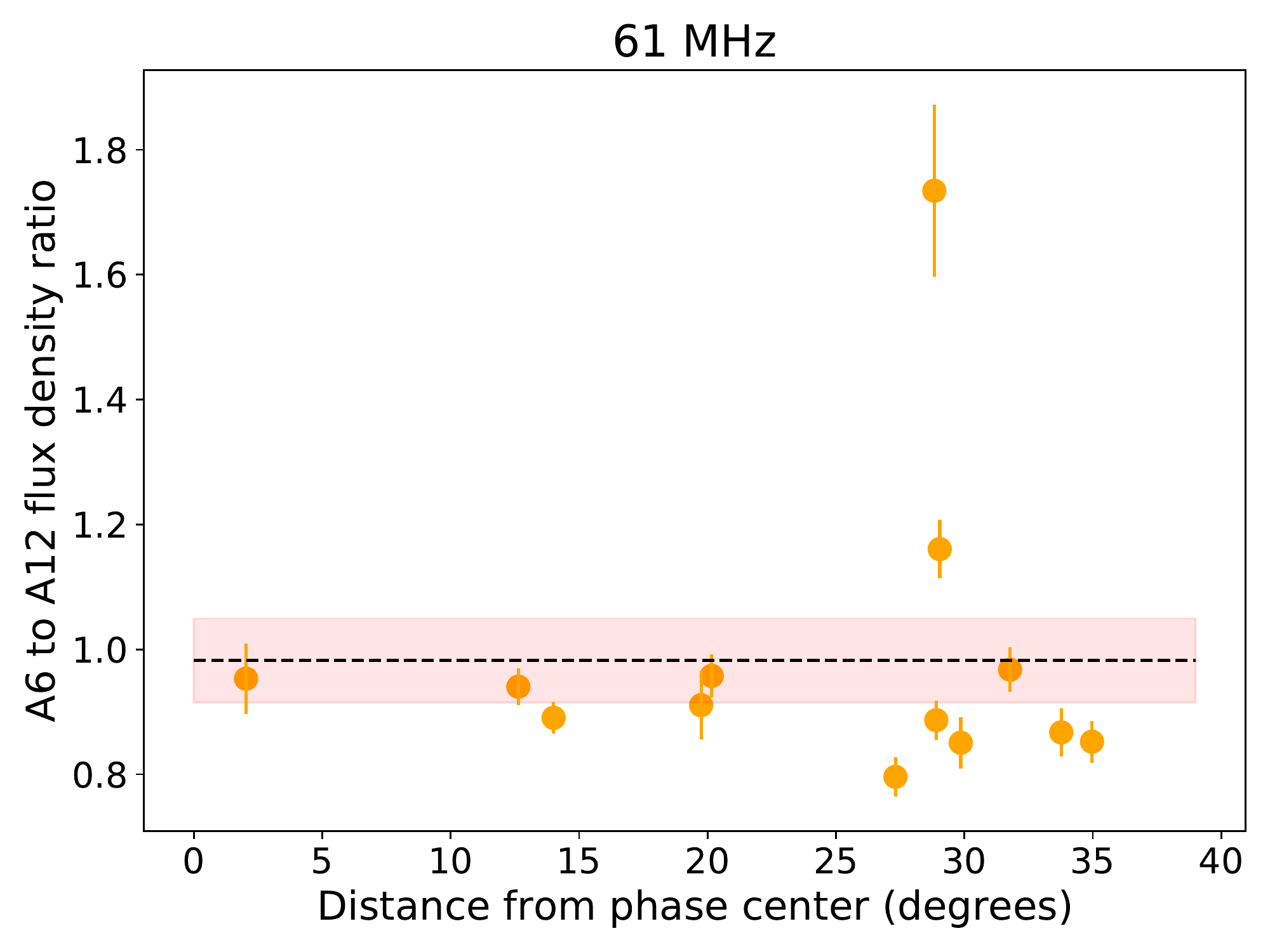}
    \caption{Source flux density ratio (8C survey and AARTFAAC-6 catalogue over A12 measured flux densities at a given frequency) across the analysis field. Plotted values are for point sources as described in the text. The black dashed line marks the mean value of the plotted data points, while the red area denotes the standard error for the sample.}
    \label{fig:fscale}
\end{figure*}

To check whether the flux scaling in the image was set correctly in the initial calibration step, we have used PyBDSF \citep{Rafferty2015} and extracted all the sources detected in a circular region with a radius of $ 38^{\circ} $ centered on zenith. The source detection settings have been set to the default values (3$\sigma$ and 5$\sigma$ for the island and peak detection thresholds respectively) and used a 2D rms map for the background ({\tt rms\_map=True}) with an {\tt rms\_box} of 30 pixels and a step of 10 pixels used to calculate the rms map.
Since A12 has a large synthesized PSF (see Table \ref{tab:noise}), and is sensitive to extended emission, we have matched our source catalogues to the 38 MHz 8C survey \citep[4.5\arcmin\ PSF;][]{Hales1995} and the 60 MHz AARTFAAC-6 survey catalog \citep[60\arcmin\ PSF;][]{Kuiack2019}. We have used a match radius of 5\arcmin\ and 17\arcmin\ at 41.7 MHz and 61 MHz respectively and did not scale the comparison catalogue fluxes due to the proximity in frequency. From the matched catalogues, we have selected the point sources as follows. At 41.7 MHz, we have selected the sources for which their A12 peak to total flux density ratio was greater than 0.5 and their size parameter in the 8C catalogue was between 0.95 and 1.2. At 61 MHz, we have selected the sources for which their A12 peak to total flux density ratio was greater than 0.85. Then, to determine the flux scale correction for each image, for the matched and selected sources we have computed the peak flux density ratio between the 8C catalogue and the 41.7 MHz A12 catalogue as well as the total flux density ratio between the AARTFAAC-6 survey catalogue and the 61 MHz A12 catalogue. The flux scale correction is taken to be the mean of the computed ratios The correction factors are $ 0.72 \pm 0.03 $ and $ 0.98 \pm 0.07 $ at 41.7 MHz and 61 MHz respectively (as shown in Figure \ref{fig:fscale}) and we use them to scale the A12 catalogue flux densities at the corresponding frequencies in the subsequent analysis.

\section{Results}
\label{sec:res}

We limit our analysis to the regions of the FoV free from Galactic emission with the highest intensity. One can model this emission and subtract it from the data, however due to the complexity of the model, we have deferred this approach for the time being and decided to use a UV taper as mentioned previously.

\begin{figure*}
	\includegraphics[width=0.45\textwidth]{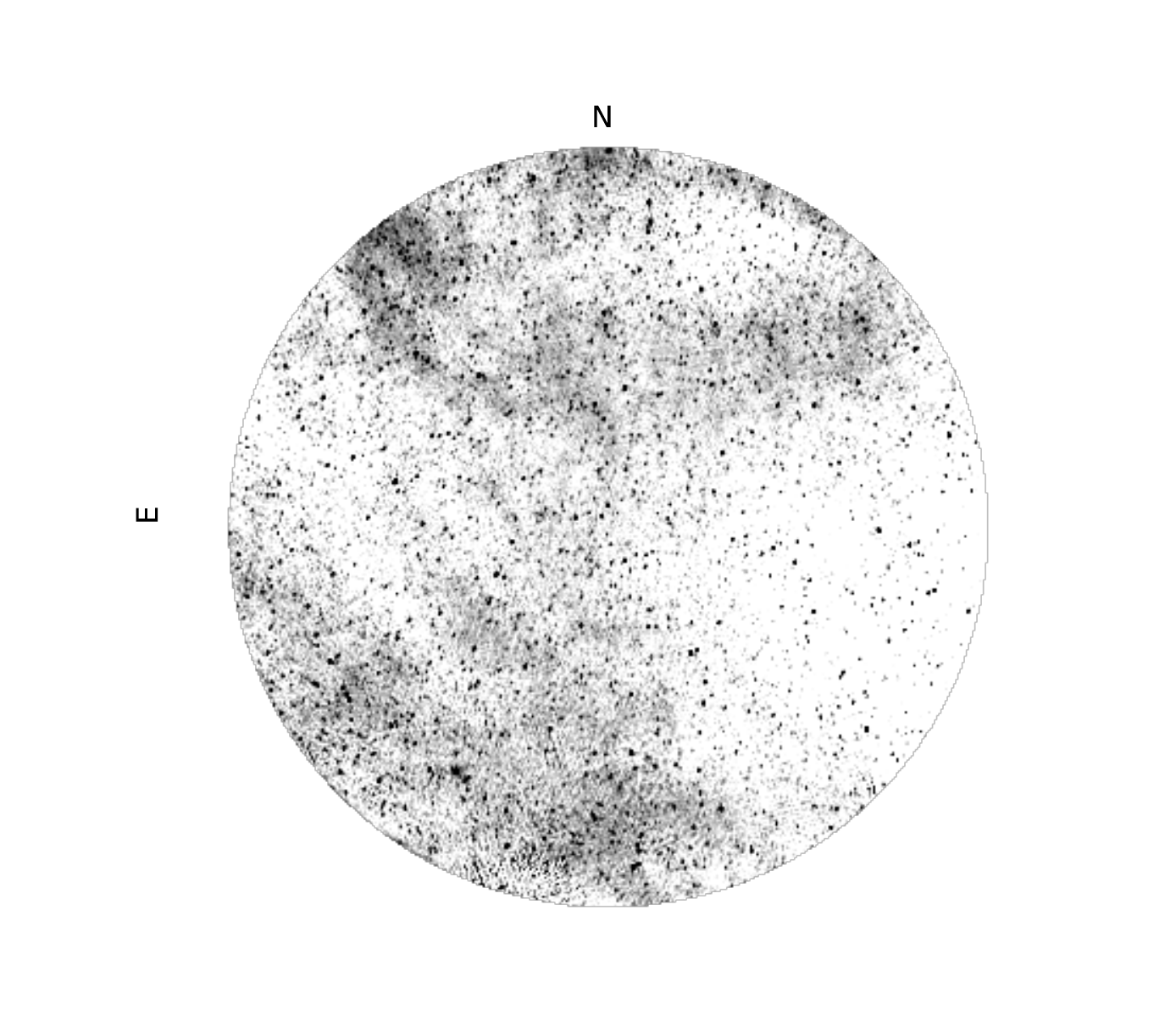}
	\includegraphics[width=0.45\textwidth]{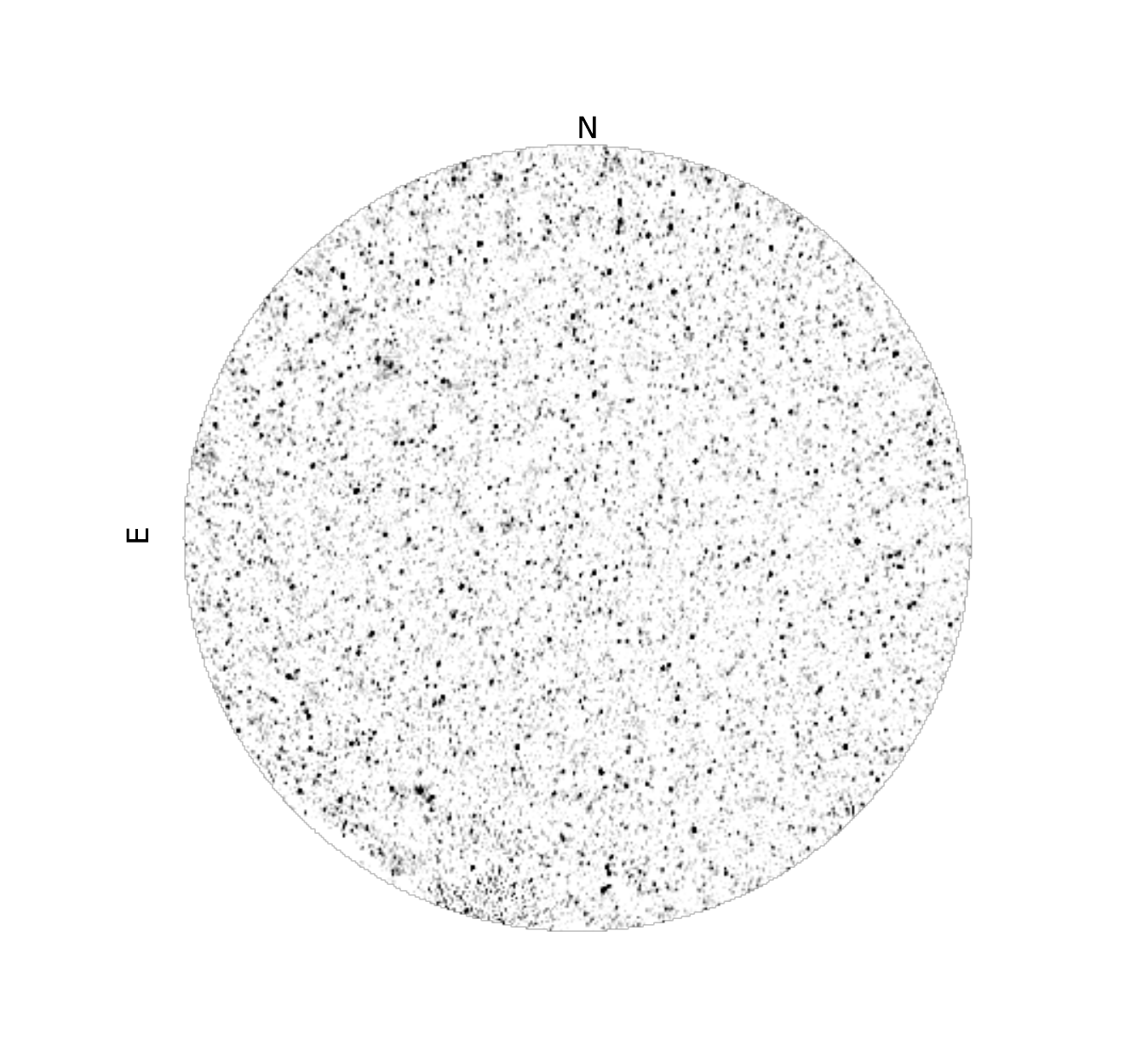} \\
	\includegraphics[width=0.45\textwidth]{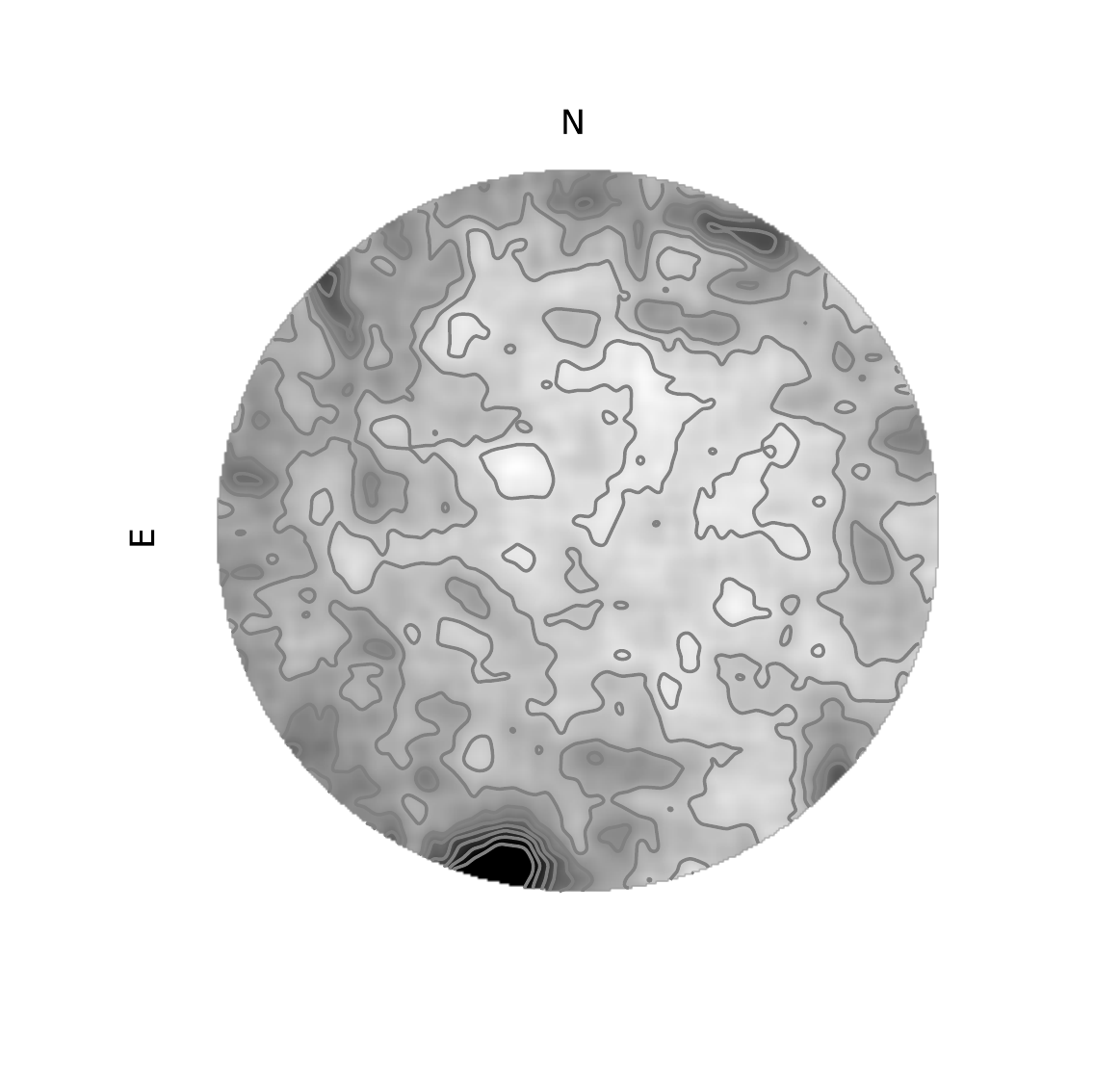}
	\includegraphics[width=0.45\textwidth]{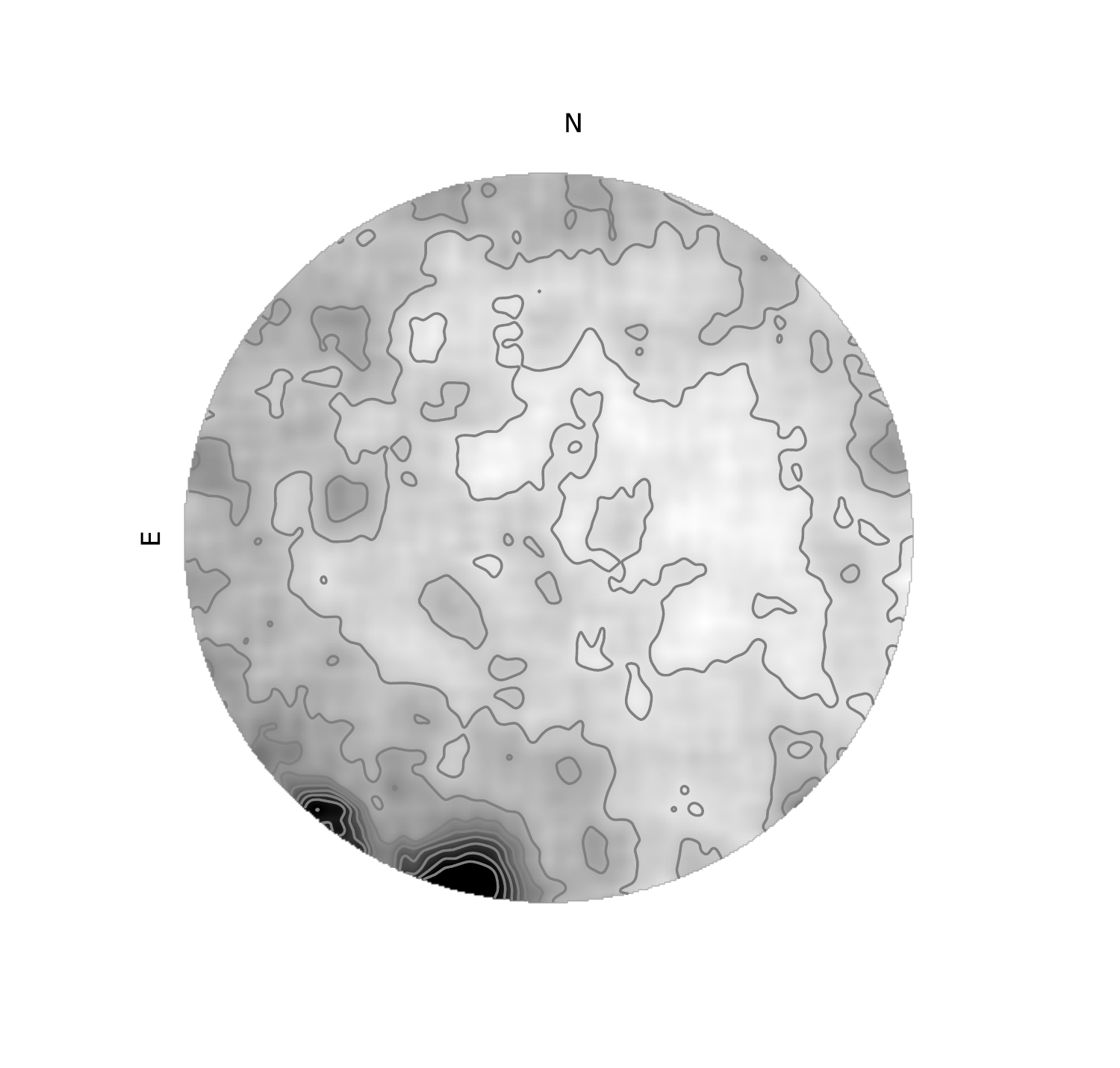}
	\includegraphics[width=0.45\textwidth]{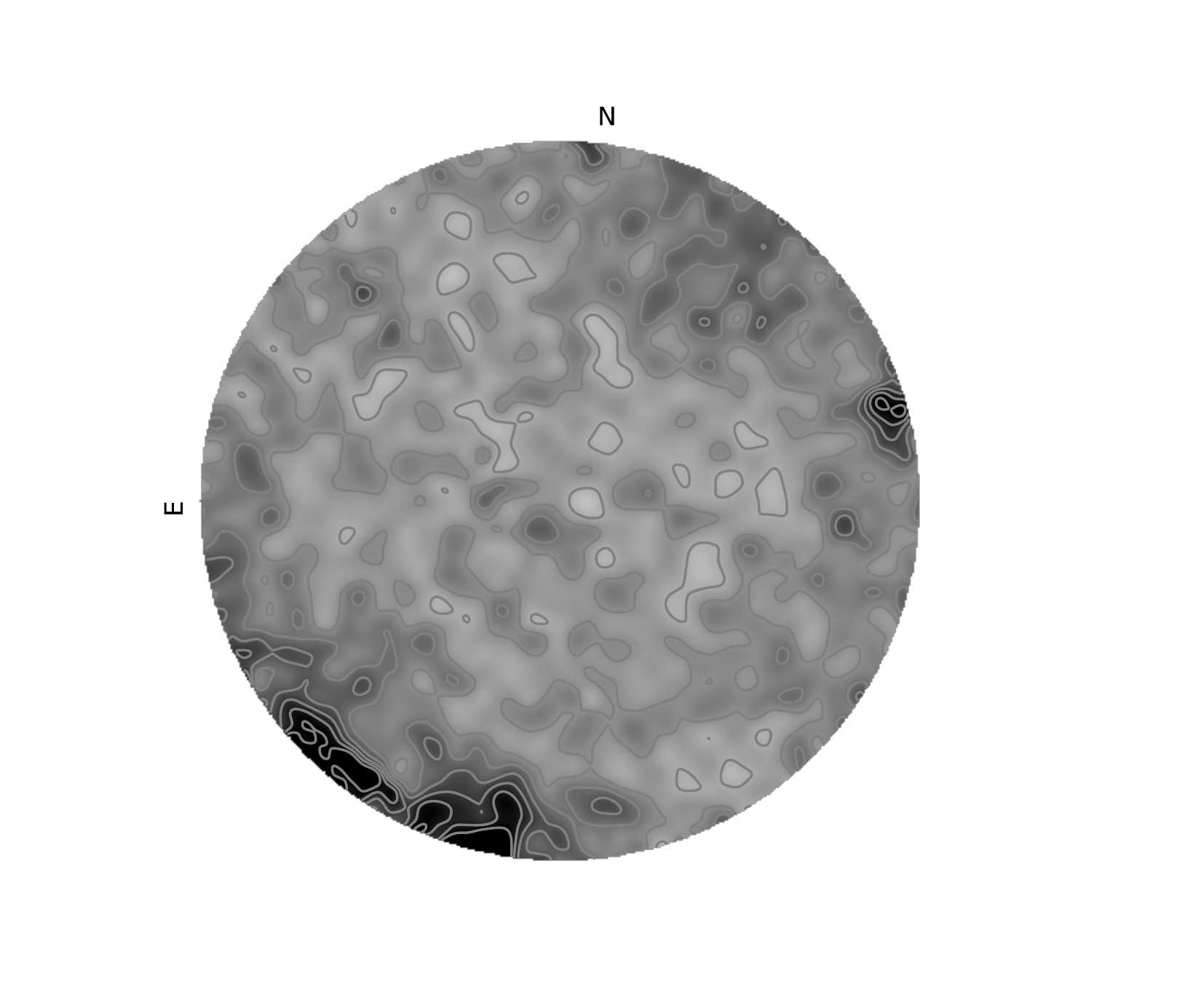}
    \caption{Image cutout of the central regions, where the analysis is performed. Top row: 61 MHz images without and with (right) UV taper. Mid row: the corresponding r.m.s. images produced by PyBDSF after source extraction is performed on the image shown in the top panel. The contours are on a linear scale going from 0.6 to 2 Jy/PSF. Bottom: UV tapered image rms at 41.7 MHz. Contours are drawn on a linear scale between 0.1 and 6 Jy/PSF. }
    \label{fig:image:panels}
\end{figure*}

The local sidereal time of the observation limits the brightest of the Galactic emission to lower elevation; we have masked the relevant region of the image and will use an area centered on zenith with a radius of 38$^{\circ}$ in our further analysis. This area is shown in the top row of Figure \ref{fig:image:panels} and the corresponding r.m.s. images (with and without UV taper) after performing source extraction with PyBDSF are given in the middle row.

\subsection{Image noise}
\label{res:noise}

For our observation duration, the images we obtain should be confusion noise limited. The theoretical confusion limit can be calculated according to \citep{vanHaarlem2013}:
\begin{equation}
\label{eq:conf}
    \sigma_{\mathrm{c}} = \dfrac{30}{10^{6}}\left(\dfrac{\theta}{1\arcsec}\right)^{1.54}\left(\dfrac{\nu}{74\,\mathrm{MHz}}\right)^{-0.7}\,\mathrm{[Jy/PSF]}
\end{equation}
\noindent where $ \theta $ is the PSF size and $ \nu $ the observing frequency. The relation is derived assuming VLSSr survey \citep{Lane2014} source counts down to a flux density limit of 0.4 Jy. 

The thermal noise for the A12 aperture array is given by:
\begin{equation}
\label{eq:thermal}
    \sigma=\dfrac{2 \eta \mathrm{k} \mathrm{T}_{\mathrm{sys}}}{\mathrm{A}_{\mathrm{eff}} 10^{-26} \sqrt{\mathrm{N}(\mathrm{N} - 1) \mathrm{p} t \Delta \nu}}\,\mathrm{[Jy/PSF]}
\end{equation}
\noindent taken from \cite{vanHaarlem2013} where $ \eta=1 $ is the assumed array efficiency, $ \mathrm{k} $ the Boltzmann constant, $ \mathrm{T}_{\mathrm{sys}} = \mathrm{T}_{\mathrm{sky}} + \mathrm{T}_{\mathrm{inst}} $ is the system temperature in degrees Kelvin, $ \mathrm{A}_{\mathrm{eff}} $ is the effective area of the LBA antenna (taking into account any mutual coupling), $ \mathrm{N} $ the total number of antennas (576 for A12), $ \mathrm{p} $ the number of recorded polarizations (4), $ t $ is the total integration time (900 seconds) and $ \Delta \nu $ is the bandwidth used (8 SBs, or 1.5 MHz in our case). We assume that the sky dominates the system temperature at low frequencies (this approximation holds best around 58 MHz), so: $ \mathrm{T}_{\mathrm{sys}} = \mathrm{T}_{\mathrm{sky}} $, and we determine the sky temperature according to: $ \mathrm{T}_\mathrm{sky} = 60 \lambda ^{2.55} $ \citep{vanHaarlem2013}, where $ \lambda $ is the observing wavelength in meters.

For the theoretical thermal noise under these assumptions, the obtained value is: $ \sigma_{\mathrm{th}} = 0.06 $ and $ 0.05 $ Jy/PSF at 41.7 and 61 MHz respectively.
The Stokes \textit{V} image noise is a good estimator of the theoretical thermal noise; we have produced Stokes \textit{V} images on various time-scales and confirmed that the Stokes \textit{V} image noise scales with the observing time as expected ($\frac{1}{\sqrt{\Delta t}}$). For the observing time we use in this analysis ($ \mathrm{t} = 15 ^{\mathrm{m}} $), the measured Stokes \textit{V} image noise is $ 0.69 $Jy/PSF and $ 0.33 $Jy/PSF at 41.7 and 61 MHz respectively. The excess over the theoretical values may be the result of flux leakage from Stokes \textit{I} to Stokes \textit{V} due to element beam model systematics.

\begin{figure*}
	\includegraphics[width=0.45\textwidth]{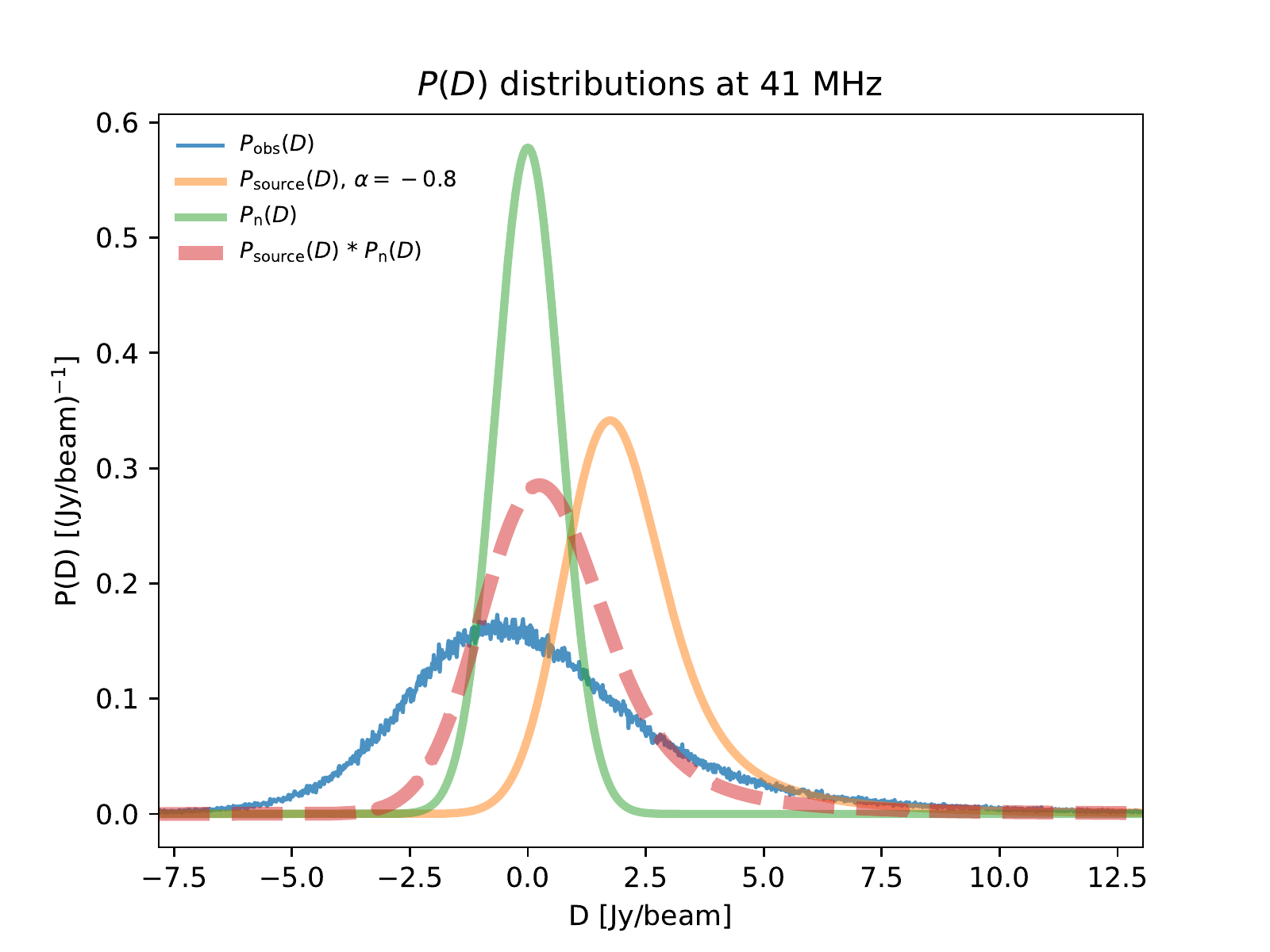}
	\includegraphics[width=0.45\textwidth]{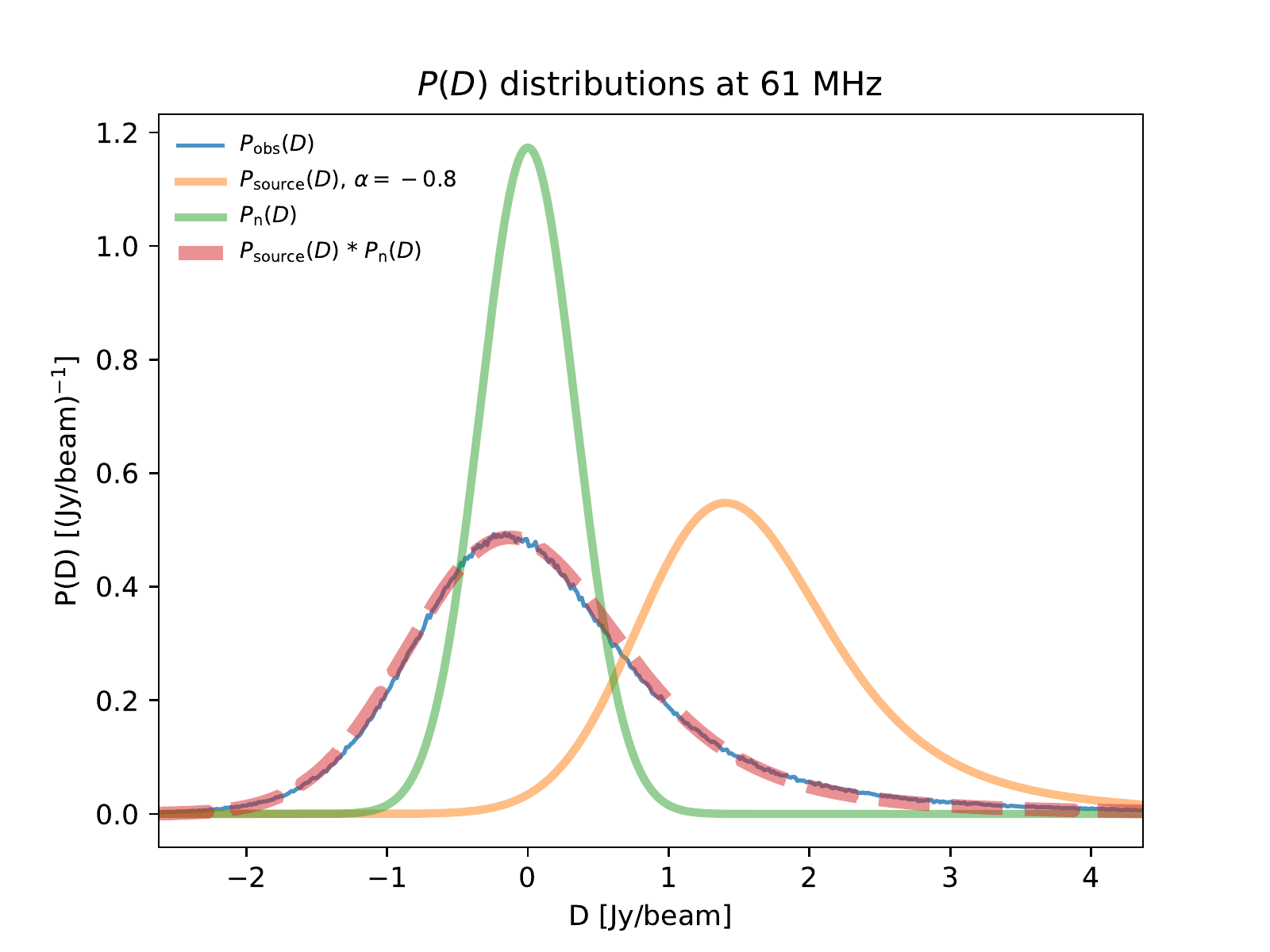}
    \caption{Left: $P(D)$ distributions at 41 MHz. The filled line Gaussian centered at zero represents the image noise probability distribution function, measured in a region of the UV tapered Stokes V image. The red dashed line represents the $P(D)$ function of the convolution between the noise and source model (orange) probability distributions. The remaining function, marked in blue, represents the smoothed probability distribution function of the pixels in the Stokes I image. Right: The same quantities calculated for the corresponding 61 MHz images. Note that the measured probability distribution matches the theoretically derived one.}
    \label{fig:image:pofd}
\end{figure*}

We use the "probability of deflection" analysis method \citep{Scheur1957} to better understand the noise properties of the images. Detailed description of the $P(\mathrm{D})$ method can be found in \cite{Vernstrom2014} and \cite{Franzen2019}, and here we present an overview of the procedure we have implemented. We convert the images to Zenithal equal area (ZEA) projection using the \textsc{Montage} package before doing the analysis, to ensure that distortions in the image plane are handled properly. For a given source count model, we compute its probability density function, i.e. the $ P_{\mathrm{source}}(D) $, taking the dirty beam into account. The source count distribution is taken to be a polynomial fit to the 154 MHz LOFAR and MWA source counts scaled to our observing frequencies \citep{Franzen2019} using a spectral index of $ -0.8 $. Then, we assume additive Gaussian (thermal) noise contribution which we estimate by measuring the standard deviation of the Stokes V images. Ideally, they should be only affected by thermal noise, assuming that the sky signal is unpolarized at the scales we sample and that there is no polarization leakage. Since the random variables describing the thermal noise and the signal produced by the source counts mentioned above are assumed to be independent, the combined probability distribution is a convolution between the probability distributions of the noise and the source model: $ P(D) = P_{\mathrm{n}}(D) * P_{\mathrm{source}}(D) $. We compare the convolution result with the measured probability density function obtained from the pixel values of the Stokes I image, $ P_\mathrm{obs}(D) $. The resulting plots are shown in Figure \ref{fig:image:pofd}. 

We derive a Stokes I image (confusion) noise of 2.6 Jy/PSF and 0.9 Jy/PSF at 41.7 and 61 MHz respectively by dividing the interquartile range with 1.349, i.e. the rms of a Gaussian distribution. At the higher frequency this value matches nicely the value estimated by convolving the SKADS source model confusion noise with the thermal noise (the width of the red dashed Gaussian in the right panel of Figure \ref{fig:image:pofd}). In the lower frequency band, there is an excess measured image noise, most likely coming from residual sidelobes of bright sources. The theoretical and measured noise values as well as detected source numbers \citep{Cohen2006} for both observing frequencies are summarized in Table \ref{tab:noise}.

\begin{table*}
	\centering
	\caption{A12 image properties. The noise is measured in Jy/PSF. Column (1) - Observing frequency, (2) - Confusion noise calculated according to the Equation \ref{eq:conf}, (3) - Confusion noise calculated based on the SKADS source models P(D) distribution. (4) - Stokes I image noise derived from the P(D) distribution analysis. (5) - Theoretical thermal noise calculated according to Equation \ref{eq:thermal}. (6) - Stokes I image noise (dashed line in Figure \ref{fig:image:pofd}) derived by convolving $ \sigma_{\mathrm{c}}^{\mathrm{P_{src}}(D)}$ and the thermal noise distribution. (7) - Image noise measured in a randomly selected Stokes \textit{V} image region. (8) - Theoretically detectable number of sources above 5$\sigma_{\mathrm{c}}^{\mathrm{P_{obs}(D)}}$. (9) - Total number of detected sources in the catalogue. (10) - PSF size}
	\label{tab:noise}
	\begin{tabular}{cccccccccc}
		\hline
		$ \nu $ [MHz] & $ \sigma_{\mathrm{c}}^{\mathrm{calc}} $ & $ \sigma_{\mathrm{c}}^{\mathrm{P_{src}(D)}} $ & $ \sigma_{\mathrm{c}}^{\mathrm{P_{obs}(D)}} $ & $ \sigma_{\mathrm{th}}^{\mathrm{calc}} $ & $ \sigma_{\mathrm{c}}^{P_{\mathrm{conv}}(D)} $ & $ \sigma_{\mathrm{V}} $ & $ \mathrm{N}_{\mathrm{th}} $ & $ \mathrm{N}_{\mathrm{det}} $ & PSF \\
		\hline
		41.7 & 2.044 & 1.239 & 2.608 & 0.059 & 1.451 & 0.686 & 450 & 147 & $24\arcmin \times 18\arcmin$\\
		61 & 0.872 & 0.766 & 0.874 & 0.047 & 0.848 & 0.330 & 1320 & 868 & $17\arcmin \times 11\arcmin $\\
		\hline
	\end{tabular}
\end{table*}

\subsection{Source counts}
\label{res:srcc}

As an initial step in deriving the source counts, we have extracted sources from the (ZEA converted) image (covering 3958 square degrees, 1.2 sr) using the \textsc{PyBDSF} source finder. We used the default parameters of the {\tt process\_image} task, hence the island and pixel detection thresholds were set to 3$\sigma$ and 5$\sigma$ respectively. We have adopted non-default values only for the {\tt rms\_box} which we have set to a size of 30 pixels. Also, we have selected to use a 2D background rms map.

We correct the total flux density entries in the catalogue by the factors derived in Section \ref{sec:flux_scale}; we do not correct for ionospheric smearing of the peak flux densities due to the size of our PSF.

Next, we bin the sources in twelve logarithmic spaced flux density bins and compute the A12 source counts using the procedure outlined in \cite{Lane2014} and \cite{Williams2016}. The raw counts per bin are weighted by the image area (expressed in steradians) associated to the sources in the corresponding bin. The computation of the areas proceeded as follows. For each source in a given bin, we have summed the number of pixels in the RMS image (output by \textsc{PyBDSF} after the source extraction) which have flux densities five times smaller than the source peak flux density. We then sum over all the sources in the bin, and knowing the pixel scale, obtain the area in steradians. We further Euclidean normalize the counts by multiplying the area corrected counts per bin by $ \mathrm{S}^{2.5} $ where $ \mathrm{S} $ is the flux density of the bin center.

\begin{figure*}
	\includegraphics[width=0.45\textwidth]{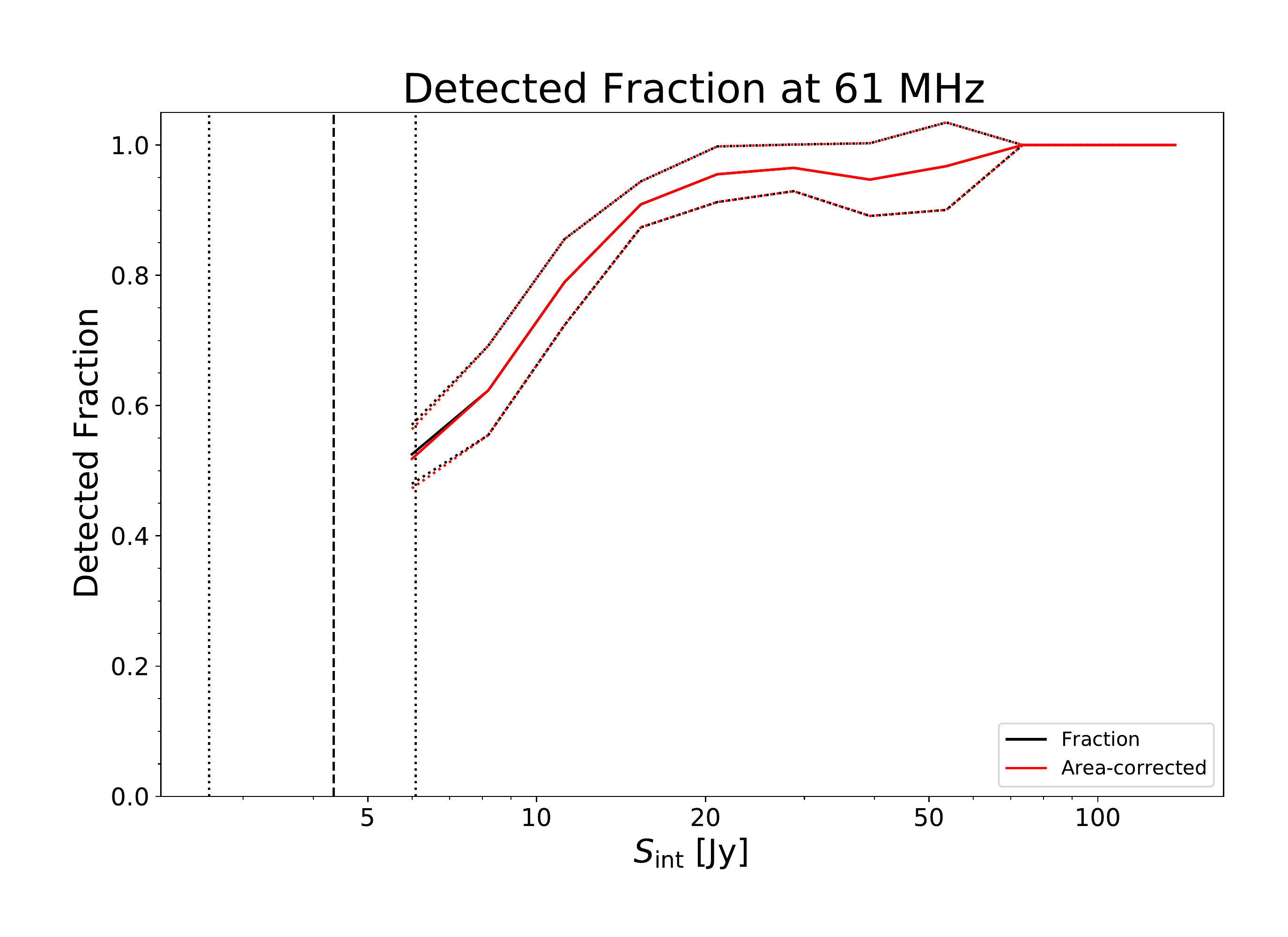}
	\includegraphics[width=0.45\textwidth]{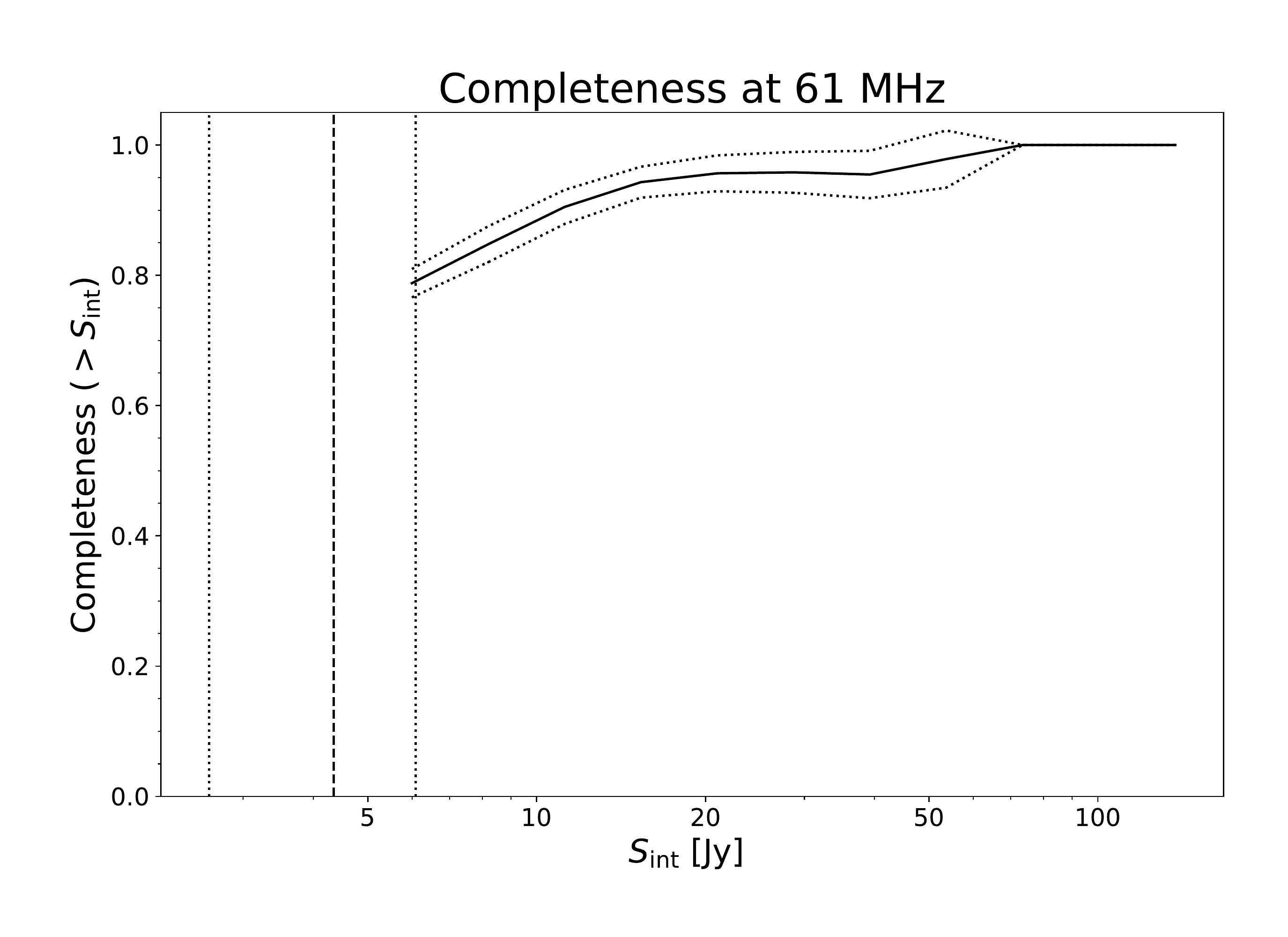} \\
	\includegraphics[width=0.45\textwidth]{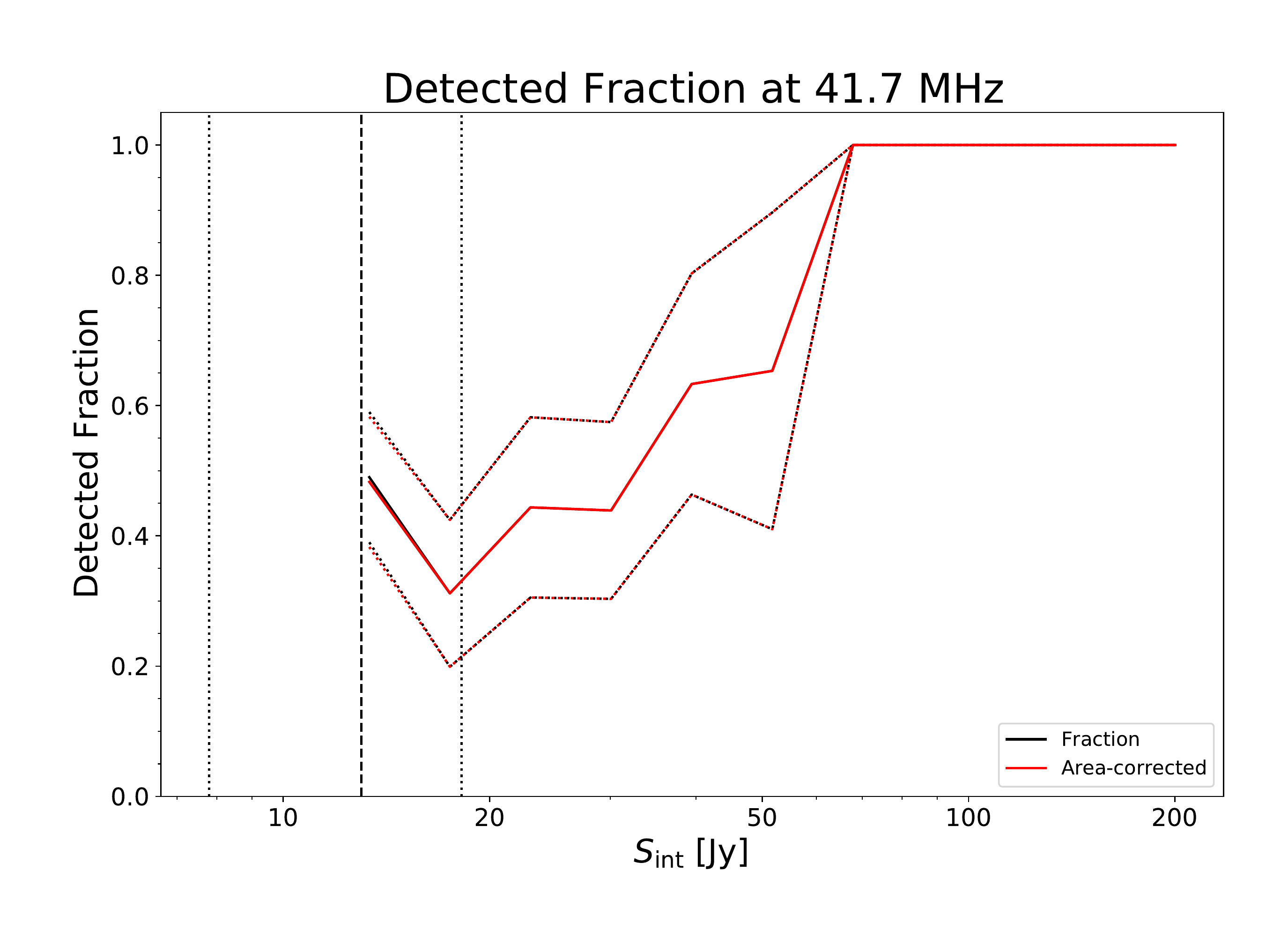}
	\includegraphics[width=0.45\textwidth]{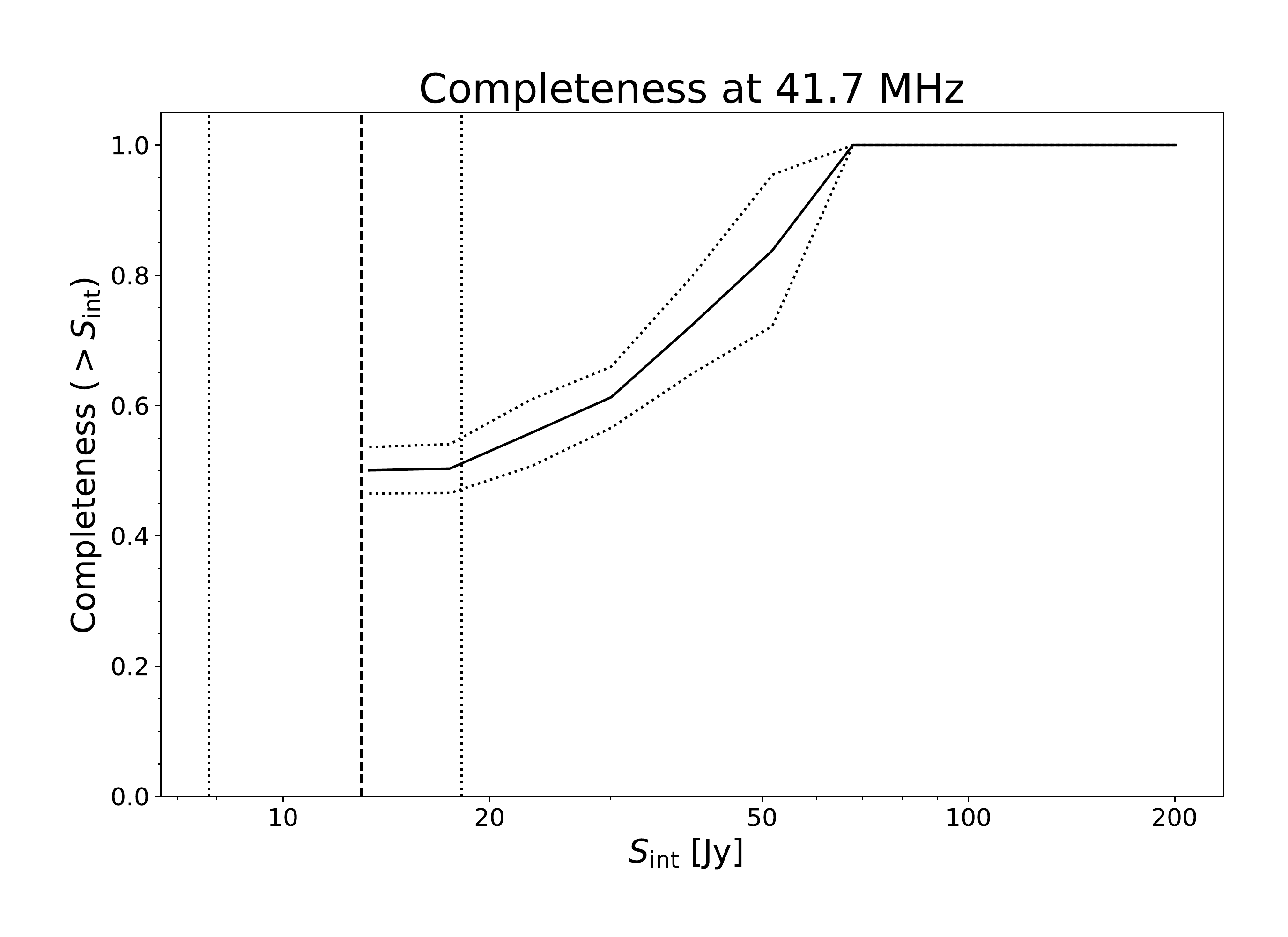}
    \caption{Top row: Source detection fraction (left) and catalogue completeness (right) at 61 MHz. The red line shows the area-corrected detection fraction. Dotted lines show its $1\sigma$ uncertainty. The vertical dashed line show the 5$\sigma$ rms noise value of the Stokes I image and the dotted vertical lines give its 2$\sigma$ bounds. Bottom row: source detection fraction and completeness at 41.7 MHz.}
    \label{fig:image:dfc}
\end{figure*}

The completeness (probability that all of the sources above a given flux density are detected) of the catalogue was obtained by performing a Monte Carlo simulation; ten images were generated by inserting 2000 sources per image at 41.7 MHz (and 4000 at 61 MHz) at random locations in the residual map output by \textsc{PyBDSF} during the source extraction mentioned above. These sources were point sources simulated as 2D Gaussians using the restoring PSF size of the original image. Their brightness was randomly drawn from a power law distribution with $ \mathrm{S}_{\mathrm{min}} = 0.2 $Jy, $ \mathrm{S}_{\mathrm{max}} = 50 $Jy and a slope of $ \alpha = -0.6 $. Out of all injected sources, 80\% were point sources. A \textsc{PyBDSF} source extraction was performed on each of the simulated images using the same parameters as the ones used during the master catalogue creation. The resulting source catalogues (from the simulated images) were matched with the input source catalogue per image using \textsc{Topcat}. These matched source catalogues were used to compute the source detection fraction, the fraction of sources actually detected per flux bin, accounting for the areas over which the source finding was performed. The completeness for a given flux density was calculated by integrating the detection fraction upwards of that flux density. The detection fraction and completeness are shown in Figure \ref{fig:image:dfc}. Our source catalogue is 97.5\% complete above 64.8 Jy at 41.7 MHz and above 50.6 Jy at 61 MHz. 

Since our PSF is large, source blending is an issue. We have derived a blend correction factor as follows. We randomly draw 2000 source fluxes between 5 and 150 Jy at 41.7 MHz (4000 source fluxes between 1 and 150 Jy at 61 MHz) from a power law distribution (as was done for the completeness correction) and inject them into the Stokes I image. We perform source extraction on these images with the same parameters as before, and obtain simulated catalogues. Source matching these catalogues to the input catalogue, we get to the blend correction, by looking at the fraction of simulated sources detected close to real sources per flux density bin \citep{Franzen2019}. We have also computed the reliability of the catalogue and its associated false detection rate and found that the associated correction is negligible, so we have not corrected the counts for it.

The derived source counts have been corrected for completeness, by multiplying them with the (area corrected) detection fraction (shown in the left panel of Figure \ref{fig:image:dfc}) and for blending by multiplying them by the derived correction factor.

\begin{figure*}
    \includegraphics[width=0.45\textwidth]{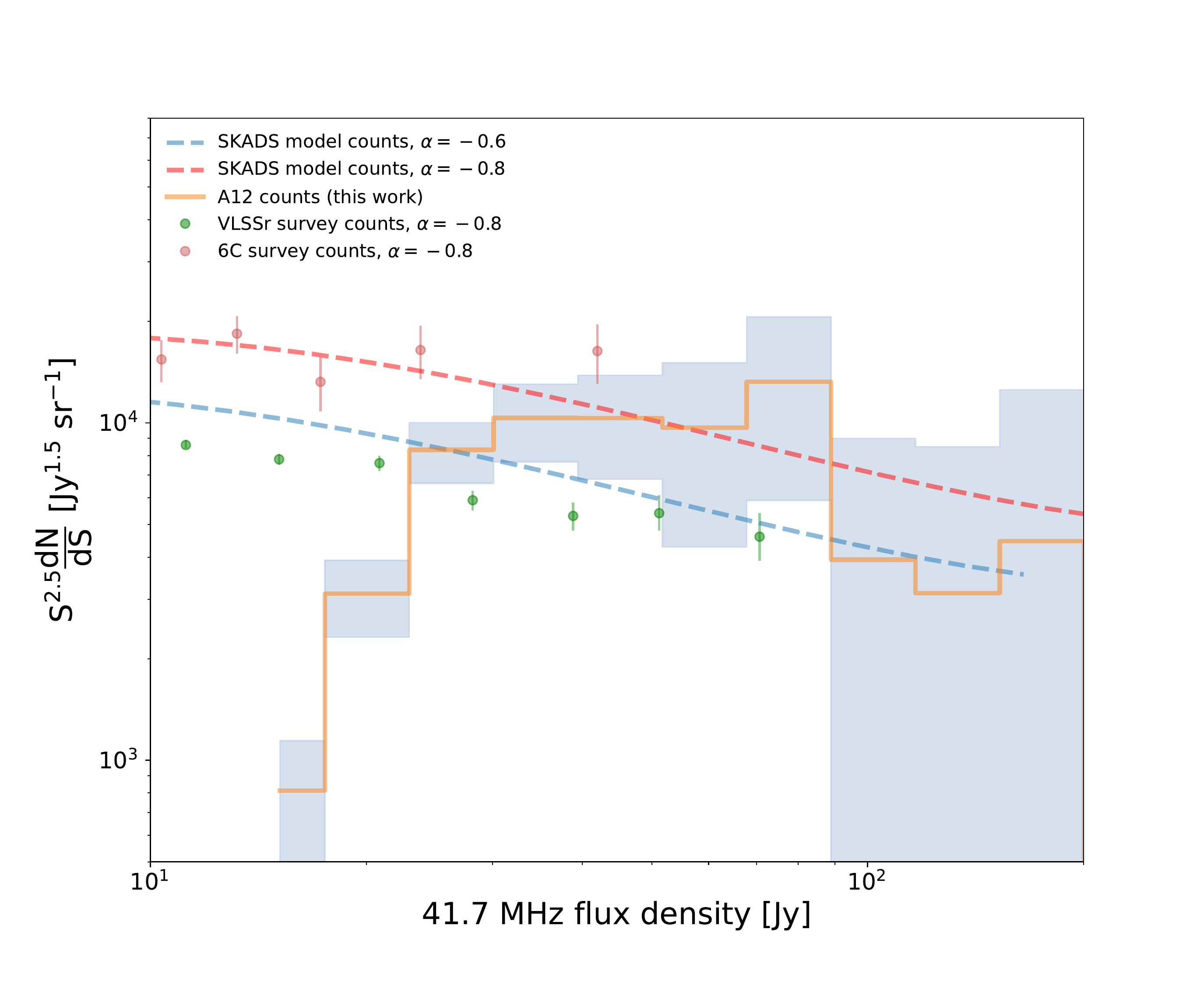}
	\includegraphics[width=0.45\textwidth]{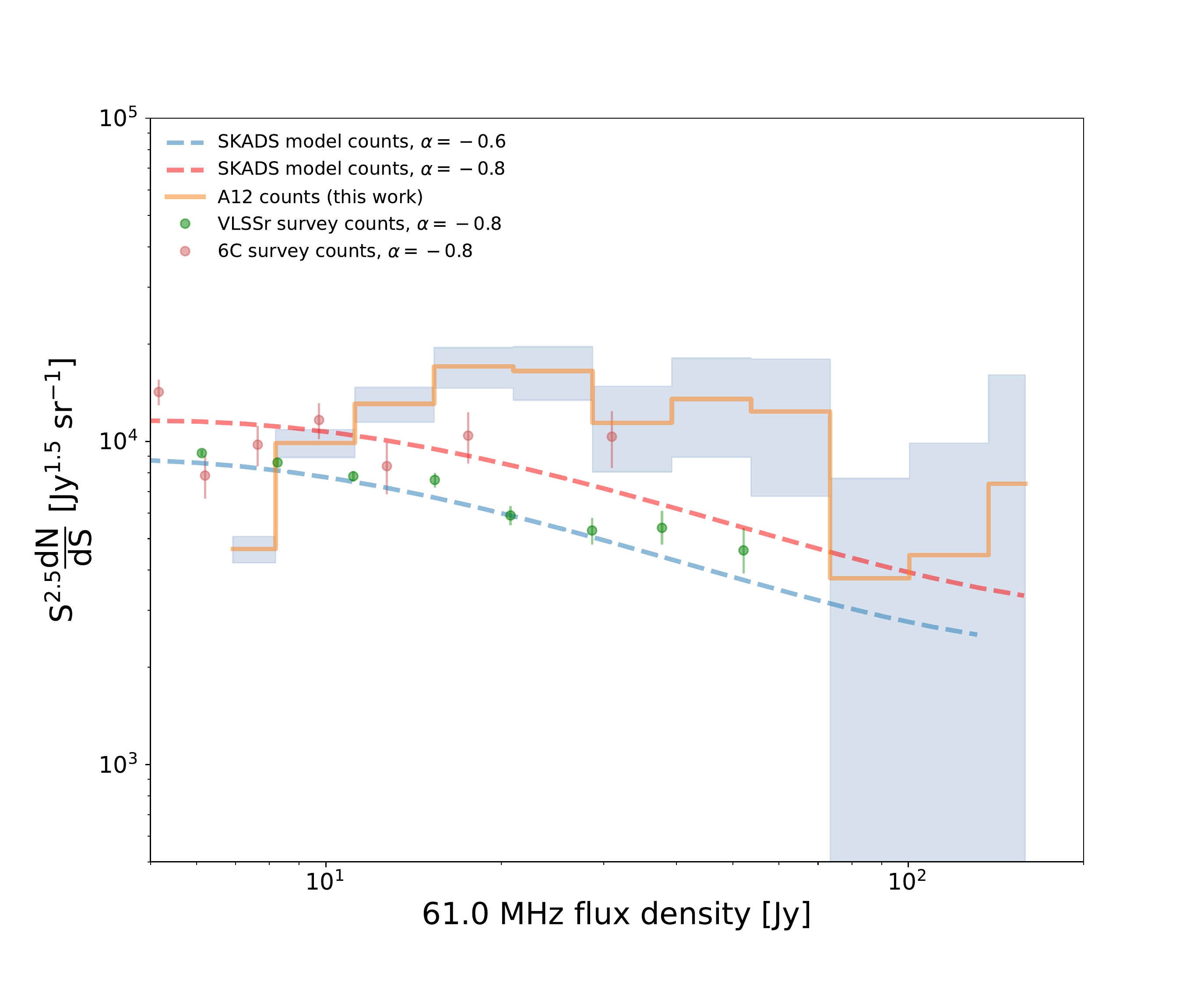}
    \caption{The derived Euclidean normalized A12 source counts (orange step line) measured at 41.7 MHz (left) and 61 MHz (right), shown with scaled 6C and VLSSr survey source counts (points). A scaled LOFAR and MWA (SKADS) source count model at the observing frequencies using a spectral index of $ -0.8 $ is also plotted for comparison. The A12 count errors are Poissonian (marked with a shaded blue region), and the flux density errors correspond to the bins used for deriving the counts. Details are given in Table \ref{tab:a12cnt}}
    \label{fig:image:counts}
\end{figure*}

 The source counts are shown in Figure \ref{fig:image:counts} along with scaled 6C survey \citep{Hales1988} and VLSSr survey derived counts as well as a SKADS model-derived source counts. Details are given in Table \ref{tab:a12cnt}.

\begin{table*}
	\centering
	\caption{A12 source counts: column (1) - flux density bin limits, (2) - central flux density, (3) - Raw source counts, (4) - Mean image area corresponding to the bin: $ <A> = 1/N \sum A $ $  $, (5) - Mean weight corresponding to the bin: $ <W> = 1/N \sum 1/A $, (6) - Completeness correction factor, (7) - Blending correction factor, (8) - Euclidean normalized source counts}
	\label{tab:a12cnt}
	\begin{tabular}{cccccccc}
		\hline
		Flux density range [Jy] & $ S_{c} $ & Raw Counts & <A> [Sr] & <W> & C.corr & B.corr & Normalized Counts [Jy$^{1.5}$sr$^{-1}$]\\
		\hline
		\multicolumn{8}{c}{61 MHz}\\
		\hline
		5.85 - 8.00 & 6.92 & 91 $ \pm $ 9 & 1.55 & 0.66 & 1.78 & 0.74 & 4644 $ \pm $ 435\\
		8.00 - 10.94 & 9.47 & 179 $ \pm $ 13 & 1.85 & 0.54 & 1.44 & 0.75 & 9891 $ \pm $ 975\\
		10.94 - 14.97 & 12.57 & 192 $ \pm $ 14 & 1.93 & 0.52 & 1.19 & 0.73 & 13069 $ \pm $ 1616\\
		14.97 - 20.47 & 17.72 & 172 $ \pm $ 13 & 1.98 & 0.50 & 1.10 & 0.76 & 17074 $ \pm $ 2446\\
		20.47 - 28.01 & 24.24 & 108 $ \pm $ 10 & 1.99 & 0.50 & 1.04 & 0.76 & 16520 $ \pm $ 3101\\
		28.01 - 38.31 & 33.16 & 50 $ \pm $ 7 & 1.99 & 0.50 & 1.00 & 0.74 & 11417 $ \pm $ 3375\\
		38.31 - 52.40 & 45.36 & 36 $ \pm $ 6 & 2.00 & 0.50 & 1.00 & 0.76 & 13525 $ \pm $ 4582\\
		52.40 - 71.68 & 62.04 & 21 $ \pm $ 5 & 2.00 & 0.50 & 1.00 & 0.75 & 12365 $ \pm $ 5598\\
		71.68 - 98.05 & 84.86 & 4 $ \pm $ 2 & 2.00 & 0.50 & 1.00 & 0.75 &  3769 $ \pm $ 3909\\
		98.05 - 134.11 & 116.08 & 3 $ \pm $ 2 & 2.00 & 0.50 & 1.00 & 0.73 & 4451 $ \pm $ 5415\\
		134.11 - 183.45 & 158.78 & 3 $ \pm $ 2 & 2.00 & 0.50 & 1.00 & 0.76 & 7391 $ \pm $ 8664\\
		\hline
		\multicolumn{8}{c}{41.7 MHz}\\
		\hline
		17.19 - 22.53 & 19.86 & 13 $ \pm $ 4 & 1.21 & 0.82 & 2.76 & 0.65 & 3174 $ \pm $ 798\\
		22.53 - 29.54 & 26.04 & 26 $ \pm $ 5 & 1.25 & 0.80 & 1.94 & 0.71 & 7126 $ \pm $ 1694\\
		526.04 - 38.73 & 34.14 & 29 $ \pm $ 5 & 1.25 & 0.80 & 1.94 & 0.63 & 10497 $ \pm $ 2685\\
		38.73 - 50.78 & 44.76 & 22 $ \pm $ 5 & 1.25 & 0.79 & 1.94 & 0.68 & 12861 $ \pm $ 3511\\
		50.78 - 66.57 & 58.68 & 23 $ \pm $ 5 & 1.25 & 0.79 & 1.00 & 0.63 & 9677 $ \pm $ 5389\\
		66.57 - 87.28 & 76.93 & 19 $ \pm $ 4 & 1.25 & 0.79 & 1.00 & 0.70 & 13243 $ \pm $ 7353\\
		87.28 - 114.43 & 100.85 & 4 $ \pm $ 2 & 1.25 & 0.79 & 1.00 & 0.66 & 3930 $ \pm $ 5064\\
		114.43 - 150.02 & 132.22 & 2 $ \pm $ 1 & 1.25 & 0.79 & 1.00 & 0.70 & 3127 $ \pm $ 5376\\
		150.02 - 196.68 & 173.35 & 2 $ \pm $ 1 & 1.25 & 0.79 & 1.00 & 0.66 & 4465 $ \pm $ 8070\\
		\hline
	\end{tabular}
\end{table*}

\subsection{Spectral index}
\label{res:spix}

\begin{figure}
	\includegraphics[width=0.5\textwidth]{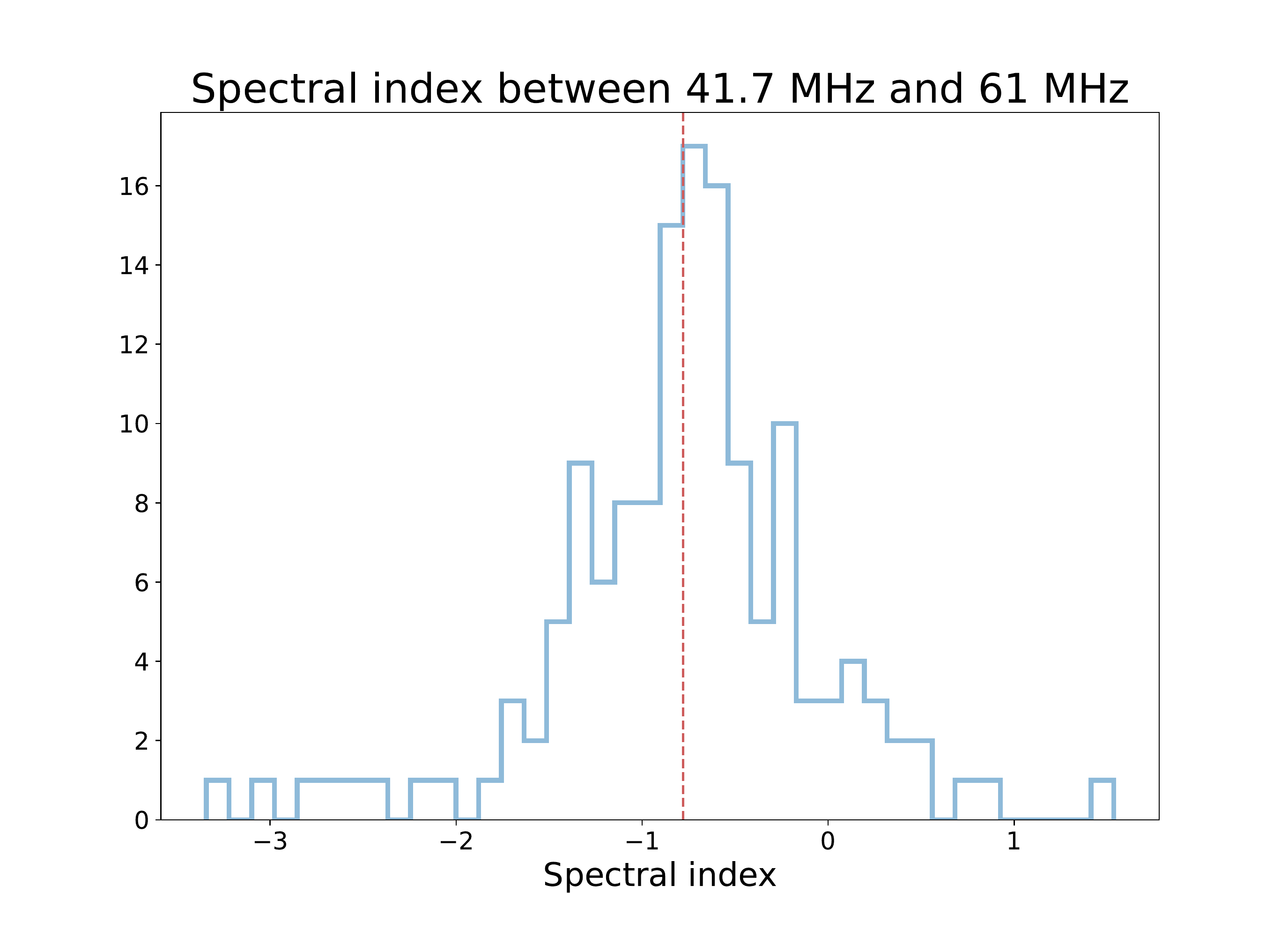}
    \caption{Spectral index distribution. The vertical dashed line represents its mean value, $\alpha_{41.7}^{61}=-0.78$.}
    \label{fig:spix}
\end{figure}

From the source catalogues at both frequencies, we have computed the two-point spectral index ($\alpha=\frac{\mathrm{log(S_{1}/S_{2})}}{\mathrm{log}(\nu_{1}/\nu_{2})}$, where $S_{1}$, $ S_{2} $ and $\nu_{1}, \nu_{2}$ are the integrated flux densities at the respective frequencies), by cross-matching the sources using a match radius of 5$\arcmin$. The resulting spectral index distribution is given in Figure \ref{fig:spix}; we find a mean spectral index value of $ \alpha_{41.7}^{61} = -0.78 $.

\section{Discussion}
\label{sec:disc}

From the image noise properties listed in Table \ref{tab:noise}, we see that the measured Stokes \textit{I} image noise at 61 MHz (estimated from the image during the P(D) analysis as well as arrived at by convolving the source model P(D) distribution with the thermal noise distribution) is in agreement with the theoretically derived confusion noise value. We conclude that A12 is confusion noise limited at 61 MHz at 0.9 Jy/PSF.
The analysis for 41.7 MHz shows that there is an excess noise component above the theoretical confusion noise limit of around 0.6 Jy/PSF most likely due to Galactic emission residuals and/or sidelobe noise from imperfect cleaning from bright sources. At this frequency, the instrument noise floor in Stokes I is 2.6 Jy/PSF. The noise floor affects the faint end of our source counts, which start at around 8$\sigma$ in the respective bands.

The A12 source counts (Figure \ref{fig:image:counts}) are in general agreement with the 6C and VLSSr survey counts, but go on to much larger flux density values. At lower flux densities the source counts show a drop off due to incompleteness. We noticed that at 61 MHz the derived source counts have an excess compared to the 6C and VLSSr survey and SKADS model counts for flux densities in the range between 20 and 80 Jy. We ascribe this to the image noise morphology (Figure \ref{fig:image:panels}) at the spatial scales and flux densities which affect the source count normalization. To correct for this systematic, we have convolved the noise map with a Gaussian kernel (100 pixel wide and having a support of 199 pixels) and used the corrected noise map for the computation of the counts shown in the right panel in Figure \ref{fig:image:counts}.

The computed spectral indices for the catalogued sources have a peaked unimodal distribution with a mean value of $ \alpha_{41.7}^{61} = -0.78 $ which is to be expected from a population of extragalactic sources, consisting predominantly of active AGN.

Following the work of \cite{Hardcastle2021} we estimate the brightness of the cosmic radio background (CRB), as reported by \cite{Fixsen2011} using the A12 source counts we derived. By integrating our source counts (corrected for completeness), we get a value of 153 K at 41.7 MHz and 133 K at 61 MHz. Both of these values are below the values predicted by the power law model proposed by \cite{Fixsen2011} (1814 K and 4909 K respectively when extrapolated to our observing frequencies) and below the values obtained if we extrapolate the values found by \cite{Hardcastle2021} at 144 MHz using a $ \mathrm{T}\sim\nu^{\beta}, \beta=-2.7 $ model (1249 K and 447 K respectively). Similarly, \cite{Subrahmanyan2013} and \cite{Dowell2018} find higher (extrapolated) values compared to what we report. 
On the other hand, \cite{Vernstrom2011} have used source counts at seven different frequencies spanning the range from 150 MHz to 8400 MHz to derive a power law dependence for the CRB, which gives a value of 18 K at 150 MHz using their derived power law index of $\beta = -2.28$. At our observing frequencies, their derived dependence gives the values of 333 K  and 140 K at 41.7 MHz and 61 MHz respectively, roughly matching our measurement at our higher observing frequency.
Our findings suggest that any extended emission on arcminute scales to which we are sensitive (for example, extended diffuse emission associated with the sources we detect but resolved out by other surveys) does not significantly contribute to the proposed background. While fainter sources detected in other (higher frequency) surveys can partially account for the difference between the CRB values we derive from our source counts and other studies, it is more likely that the cause can be ascribed to residual systematics (especially pertaining to the lowest frequencies we probe) coupled with a possible low frequency turnover in the power law fits to existing source counts.

\section{Conclusions and future prospects}
\label{sec:conc}

The AARTFAAC-12 aperture array is a very capable instrument, delivering confusion limited images which we have analyzed in this work and shown that they are applicable in various science cases. The instrument is particularly suitable for transient science as well as imaging large scale diffuse emission, necessary for updating existing sky models at the lowest radio frequencies. In this work we have:

\begin{itemize}
    \item Described in detail the procedure used to calibrate and image A12 data.
    \item Characterized and validated the images, using P(D) analysis to derive noise parameters. This represents a crucial precursor for science exploitation. A12 is confusion limited at around 0.9 Jy at 61 MHz.
    \item Derived source counts at 41.7 and 61 MHz, which are in agreement with previous studies and which showcase the ability of the instrument to perform large scale surveys on a short timescale.
    \item Estimated the spectral index between our observing bands and used the source counts to characterize any excess radio emission at the frequencies observed.
\end{itemize}

The A12 calibration and imaging pipeline is modular and it will be modified to speed up processing, leading to a near-real time transient detection pipeline with an imaging cadence of around one second. The instrument will evolve as the LOFAR telescope is upgraded to its more capable successor, LOFAR2.0.

\section*{Data availability}

The data underlying this article will be shared on reasonable request to the corresponding author.

\section*{Acknowledgements}

The authors would like to thank the anonymous referee for the constructive comments which have helped to improve this work.
AARTFAAC development and construction was funded by the ERC under the Advanced Investigator grant no. 247295 awarded to Prof. Ralph Wijers, University of Amsterdam.
We thank The Netherlands Institute for Radio Astronomy (ASTRON) for support provided in carrying out our observations. AARTFAAC is maintained and operated jointly by ASTRON and the University of Amsterdam.
WLW also acknowledges support from the CAS-NWO programme for radio astronomy with project number 629.001.024, which is financed by the Netherlands Organisation for Scientific Research (NWO).
BKG acknowledges the financial support from the European Research Council (ERC) under the European Union's Horizon 2020 research and innovation programme (Grant agreement No. 884760, "CoDEX") and the National Science Foundation through an award for HERA (AST-1836019).
We would also like to thank the LOFAR telescope operators and science support group for their assistance in obtaining the data used in this work.
We use data obtained from LOFAR, the Low Frequency Array designed and constructed by ASTRON, which has facilities in several countries, that are owned by various parties (each with their own funding sources), and that are collectively operated by the International LOFAR Telescope (ILT) foundation under a joint scientific policy.
This work was carried out on the Dutch national e-infrastructure with the support of SURF Cooperative (Grant No. EINF-245).
This research made use of Montage. It is funded by the National Science Foundation under Grant Number ACI-1440620, and was previously funded by the National Aeronautics and Space Administration's Earth Science Technology Office, Computation Technologies Project, under Cooperative Agreement Number NCC5-626 between NASA and the California Institute of Technology.
This research made use of {\tt Astropy}\footnote{\url{http://www.astropy.org}} a community-developed core Python package for Astronomy \citep{astropy:2013, astropy:2018}, as well as: {\tt NumPy} \citep{vanderWalt2011}, {\tt SciPy} \citep{Virtanen2019} and {\tt Matplotlib} \citep{Hunter2007}. Accordingly, we would like to thank the scientific software development community, without whom this work would not be possible.


\bibliographystyle{mnras}
\bibliography{A12_scounts} 







\bsp	
\label{lastpage}
\end{document}